\documentclass[pre,aps,twocolumn,epsfig,rotate,superscriptaddress,showpacs]{revtex4}
\usepackage{epsfig}
\usepackage{graphicx}
\usepackage{amsmath}

\newcommand{\beq}{\begin{eqnarray}}
\newcommand{\eeq}{\end{eqnarray}}
\begin{document}
\title{Phase-space characterization of complexity
in quantum many-body dynamics}
\author{Vinitha Balachandran}
\affiliation{Department of Physics and Center for
Computational Science and Engineering,
\\ National University of Singapore, Singapore 117542}
\author{Giuliano Benenti}
\affiliation{CNISM, CNR-INFM \& Center for Nonlinear and Complex Systems,
Universit\`a degli Studi dell'Insubria, Via Valleggio 11, 22100 Como, Italy}
\affiliation{Istituto Nazionale di Fisica Nucleare, Sezione di Milano,
via Celoria 16, 20133 Milano, Italy}
\author{Giulio Casati}
\affiliation{CNISM, CNR-INFM \& Center for Nonlinear and Complex Systems,
Universit\`a degli Studi dell'Insubria, Via Valleggio 11, 22100 Como, Italy}
\affiliation{Istituto Nazionale di Fisica Nucleare, Sezione di Milano,
via Celoria 16, 20133 Milano, Italy}
\affiliation{Centre for Quantum Technologies,
National University of Singapore, Singapore 117543}
\author{Jiangbin Gong}
\affiliation{Department of Physics and Center for
Computational Science and Engineering,
\\ National University of Singapore, Singapore 117542}
\affiliation{NUS Graduate School for Integrative Sciences and
Engineering, Singapore 117597}
\begin{abstract}
We propose a phase-space Wigner harmonics
entropy measure for many-body quantum dynamical complexity.
This measure, which reduces to the well known measure of complexity
in classical systems and which is valid for both pure and mixed states
in single-particle and many-body systems, takes into account
the combined role of chaos and entanglement
in the realm of quantum mechanics.
The effectiveness of the measure is
illustrated in the example of the Ising chain in a homogeneous tilted
magnetic field.
We provide numerical evidence that the multipartite entanglement generation
leads to a linear increase of entropy until
saturation in both integrable and chaotic regimes, so that
in both cases the number of harmonics of the Wigner function
grows exponentially with time.
The entropy growth rate can be used to detect quantum phase transitions.
The proposed entropy measure can also distinguish between
integrable and chaotic many-body dynamics by means of the size
of long term fluctuations which become smaller when quantum chaos
sets in.
\end{abstract}
\date{\today}
\maketitle

\section{introduction}

Understanding the dynamics of quantum systems is a challenging task
of immense importance in a variety of fields including condensed
matter physics and quantum information science.
Quantum dynamical complexity refers to the lack of a simple description
of the evolution of a quantum system.
From a computational perspective, it implies
the inevitable loss of predictability of system evolution using
classical simulation. In many-body interacting quantum systems,
complexity can be attributed to non-integrability
or to the tensor product structure of the Hilbert space.
Hence, quantum chaos and entanglement have deep implications
in characterizing quantum many-body dynamical complexity.

In classical physics,
it is very well known that there exists a direct correlation between
chaos and complexity.
  Classically chaotic systems are characterized
 by exponentially diverging nearby trajectories, with a rate
 determined by the Lyapunov exponent. Complexity then arises
from the fact that the orbits of such deterministic
systems are completely random and unpredictable with
positive algorithmic complexity~\cite{Ford}.
In quantum mechanics,
trajectories in standard treatments are forbidden by the Heisenberg uncertainty principle
and therefore the above notion of
complexity cannot be directly translated to quantum systems.

However, the phase-space approach can be equally used for both
classical and quantum mechanics.
In the context of classical systems, it has been shown
that the second moment of the Fourier components of the classical
distribution function grows linearly for an integrable system
while it grows
exponentially for a chaotic system, with a rate determined by the
Lyapunov exponent characterizing the local exponential instability.
Thus, the growth rate of the second moment of Fourier components
(harmonics) is a good measure of the complexity of classical
dynamics~\cite{Gong03}.
In a similar way, for single-particle quantum systems the second moment of
harmonics of the Wigner distribution function of a quantum state,
pure or mixed, is a measure of quantum complexity~\cite{Gong03,Casati08}.
Note that in quantum systems with few degrees of freedom an exponential growth of
the number of harmonics is possible only up to the Ehrenfest
time scale, after which the growth is at most linear~\cite{Casati08}.
Moreover, the number of harmonics of the Wigner function can
be used to detect, in the time domain, the crossover from
integrability to chaos~\cite{Benenti09}.

For quantum many-body systems the situation is more complicated.
First note that quantum dynamical entropies, which generalize the
Komologrov-Sinai entropy to quantum dynamical systems, can be
positive even for integrable dynamics~\cite{Adv}. This behavior may
appear, at least at first sight, somehow surprising since in
classical dynamics positive Komologorov-Sinai entropy implies chaos.
Another interesting property is that, as shown in
Ref.~\cite{Prosen07}, the rank of the matrix product operator
representation of the pure quantum states in the time-dependent
density-matrix renormalization group, typically grows exponentially
even for integrable system with finite number of particles. This
inefficiency of the classical simulation of many-body quantum
dynamics can be attributed to entanglement and is consistent with
the linear growth of the entanglement block entropy for integrable
spin chains~\cite{Calabrese}.

Several very interesting definitions of quantum complexity have
been proposed, e.g., see Ref.~\cite{Adv} and references therein.
On the other hand, to the best of our knowledge none of them
satisfies all the following requirements,
which a notion of complexity should possess in order to be both
meaningful and practically useful:
\begin{itemize}
\item[(i)] To provide a unified description
of both one- and many-body dynamics;
\item[(ii)] To reproduce at the classical limit
the well-known notion of classical complexity
based on the local exponential instability of chaotic dynamics;
\item[(iii)] To be applicable to both pure and mixed states;
\item [(iv)] To be practically useful, that is, convenient for
numerical investigations.
\end{itemize}
The purpose of the present paper is to propose a
notion of complexity that fulfills the above criteria.
By extending previous investigations~\cite{Gong03,Casati08,Benenti09}
to many-body quantum dynamics,
we propose the number of harmonics of the Wigner function as
a suitable measure of complexity of a quantum state.
Indeed, as the phase-space formulation of quantum dynamics
can be directly generalized to many-body systems,
the harmonics of the Wigner function seem to be very
promising in quantifying the complexity of many-body quantum
systems as well.
Hence, in this paper,
we introduce a Wigner harmonics entropy measure of complexity
and then illustrate its usefulness
by means of numerical simulations carried on a
paradigmatic spin-chain  model, the Ising chain in a tilted
magnetic field.
We will show that the entropy grows linearly until
saturation in both integrable and chaotic regimes, so that
in both cases the number of harmonics of the Wigner function
grows exponentially with time. We will provide numerical evidence that
this growth must be attributed to multipartite entanglement generation.
Our results demonstrate that the growth rate can be used also
to detect quantum phase transitions.
Finally, the proposed entropy measure can also distinguish between
integrable and chaotic many-body dynamics, by means of the size
of long term fluctuations, which become smaller when a transition
to chaos occurs.

This paper is organized as follows. In Sec.~\ref{sec:harmonics}, we define
our phase-space quantum complexity measure, based on the harmonics
of the Wigner function. The working of such measure is illustrated in
the dynamics of a many-body spin-chain model, introduced in Sec.~\ref{sec:model}
and investigated in detail in Sec.~\ref{sec:numerics}.
Finally, our conclusions are drawn in Sec.~\ref{sec:conclusions}.

\section{Harmonics of the Wigner function}

\label{sec:harmonics}

The phase-space representation of quantum mechanics is a
very enlightening approach as it allows a direct comparison
between quantum and classical dynamics. In particular,
the complexity of a quantum state or of a classical distribution
function can be measured by the richness of their phase space
structure.

In the quantum case, the phase-space approach to complexity is
particularly convenient for systems whose Hamiltonian can be
written in terms of a set of bosonic creation-annihilation
operators:
$$
 \hat{H}(\hat{a}_{1}^{\dag},...,\hat{a}_{N}^{\dag},
\hat{a}_{1},...,\hat{a}_{N};t)
  \equiv
\hat{H}^{(0)}(\hat{n}_{1},...,\hat{n}_{N})
$$
\begin{equation}\label{manyha}
+\hat{H}^{(1)}(\hat{a}_{1}^{\dag},...,\hat{a}_{N}^{\dag},
\hat{a}_{1},...,\hat{a}_{N};t),
\end{equation}
with
$[\hat{a}_{i},\hat{a}_{j}]=[\hat{a}_{i}^{\dag},\hat{a}_{j}^{\dag}]=0$,
$[\hat{a}_{i}^{\dag},\hat{a}_{j}]=\delta_{ij}$, and
the number operators $\hat{n}_i= \hat{a}_i^\dag \hat{a}_i$.

We will use the method of $c$-number ${\boldsymbol\alpha}$-phase space borrowed
from quantum optics (see for example Ref.~\cite{Phase}).
The Wigner function $W({\boldsymbol\alpha},{\boldsymbol\alpha}^{*};t)$
of a state $\hat{\rho}(t)$ is defined by
$$
W({\boldsymbol\alpha},{\boldsymbol\alpha}^{*};t)=\frac{1}{\pi^{2N}
\hbar^N} \int d^{2}{\boldsymbol\eta}
\exp{\left(\frac{{\boldsymbol\eta}^*\cdot{\boldsymbol\alpha}}{\sqrt{\hbar}}
-\frac{{\boldsymbol\eta}
\cdot{\boldsymbol\alpha}^*}{\sqrt{\hbar}}\right)}
$$
\begin{equation}
\times \text{Tr}[
\hat{\rho}(t)\hat{D}({\boldsymbol\eta})],\label{wignerm}
\end{equation}
where ${\boldsymbol\eta}=(\eta_1,...,\eta_N)$ and
${\boldsymbol\alpha}=(\alpha_1,...,\alpha_N)$ are $N$-dimensional
complex variables, the integration runs over the complex
$\eta_i$-planes for $i=1,...,N$, the displacement operator
\begin{equation}
\hat{D}\left({\boldsymbol\eta}\right)=
\exp\left[\sum_{i=1}^{N}{\left(\eta_i\hat{a_{i}}^
{\dagger}-\eta_i^{*}\hat{a_{i}}\right)}\right],
\end{equation}
and the coherent states
\begin{equation}
|{\boldsymbol\alpha}\rangle=
|\alpha_{1}\alpha_{2}...  \alpha_{N}\rangle=
\hat{D}\left(\frac{{\boldsymbol\alpha}}{\sqrt{\hbar}}\right)|00....0\rangle,
\end{equation}
with $|\alpha_i\rangle$ being eigenstate of the annihilation operator
$\hat{a}_i$, i.e., $\hat{a}_i|\alpha_i\rangle= \frac{\alpha_i}{\sqrt{\hbar}}
|\alpha_i\rangle$,
and $|00...0\rangle$ being the vacuum state.
We define the harmonic's amplitudes $W_{\bf m}({\bf I};t)$ of the
Wigner function by the $N$-dimensional Fourier expansion
\begin{equation}\label{eq:calW}
W({\boldsymbol\alpha},{\boldsymbol\alpha}^{*};t)=
\frac{1}{\pi^{N}} \sum_{\bf m} W_{\bf m}({\bf I};t)
e^{i{\bf m}\cdot {\boldsymbol\theta}},
\end{equation}
where ${\bf m}, {\bf I},\boldsymbol\theta$ are $N$-dimensional
vectors, whose components $I_k\ge 0$, $0\le \theta_k< 2\pi$ are
defined by the relations $\alpha_k=\sqrt{I_k} e^{-i\theta_k}$,
$k=1,...,N$. Here $I_k$ and $\theta_k$ can be regarded as our quantum phase space
variables, analogous to the action and angle variables in the classical phase space. Note that $W_{-{\bf m}}=W_{\bf m}^{*}$. The Wigner
function's normalization condition $\int d^{2}  {\boldsymbol\alpha}
W({\boldsymbol\alpha},{\boldsymbol\alpha}^{*};t)=1$ simply implies
that $\int d{\bf I} W_{{\bf 0}}({\bf I};t)=1$, while there are no
restrictions on $W_{\bf m}$ when ${\bf m}\ne {\bf 0}$.

In Refs.~\cite{Casati08,Benenti09}, the number of harmonics of
the Wigner function was estimated by
$\sqrt{\langle {\bf m}^2 \rangle_t}$, with
$\langle {\bf m}^2 \rangle_t$ being the second moment of the
harmonics distribution:
\begin{equation}
\langle {\bf m}^2 \rangle_t
={\sum_{\bf m}}
{\bf m}^2 \mathcal{W}_{\bf m}(t),
\end{equation}
where
\begin{equation}
\mathcal{W}_{\bf m}(t)\equiv
\frac{\int d{\bf I}
|W_{\bf m}({\bf I};t)|^2}{\sum_{\bf m} \int d{\bf I}
|W_{\bf m}({\bf I};t)|^2}.
\label{eq:mathcalW}
\end{equation}
The harmonics distribution $\mathcal{W}_{\bf m}$ is normalized,
$\sum_{\bf m} \mathcal{W}_{\bf m}=1$.
For one-body systems, the second moment $\langle {\bf m}^2 \rangle_t$
provides a reliable estimate of the number of harmonics in a generic
chaotic case~\cite{Casati08} and is able to distinguish, in the semiclassical
region, between integrable and
chaotic regimes~\cite{Benenti09}.
In the first case, $\sqrt{\langle {\bf m}^2 \rangle_t}$
grows linearly in time, in the latter exponentially. On the other hand,
we expect that the number of harmonics always
captures the complexity of motion, including the case of many-body systems
without classical analogue.
Hence, we propose as a complexity measure the entropy
\begin{equation}\label{ent}
    {S}(t)=-{\sum_{m_1,...,m_N\ge 0}} \mathcal{W}_{\bf m}(t)
    \ln [\mathcal{W}_{\bf m}(t)],
\end{equation}
where the sum over ${\bf m}$ is limited to $m_1,...,m_N\ge 0$ since
harmonics $\mathcal{W}_{\bf m}$ and $\mathcal{W}_{-{\bf m}}$ are not
independent but trivially related by the relation
$\mathcal{W}_{\bf m}=\mathcal{W}_{-{\bf m}}$ (note that the same limitation
must now be taken in Eq.~(\ref{eq:mathcalW}) in order to properly
normalize the distribution $\mathcal{W}_{\bf m}$).
The number of harmonics of a generic state $\hat{\rho}(t)$ can
therefore be measured by $\exp[{S}(t)]$.
For the models discussed in this paper both the second moment
$\langle {\bf m}^2 \rangle $ and the entropy $S$ provide qualitatively
the same results \cite{footnote}.

The main computational advantage of the above
$c$-number ${\boldsymbol\alpha}$-phase space approach is that
the Wigner function's harmonics $\mathcal{W}_{\bf m}$ can be
computed very conveniently from the density matrix written in the
basis of the eigenvectors $|{\bf n}\rangle=|n_1...n_N\rangle$ of the unperturbed
Hamiltonian $\hat{H}^{(0)}$.
Indeed, using the well-known
matrix elements of the displacement operator~\cite{Schwinger},
\begin{equation}\label{D}
\langle n_i+m_i |{\hat
D}(\eta_i)|n_i\rangle={\sqrt\frac{n_i!}{(n_i+m_i)!}}\,\eta_i^{m_i}\,
e^{-\frac{1}{2}|\eta_i|^2}L_{n_i}^{m_i} (|\eta_i|^2)\,,
\end{equation}
($n_i,m_i\geq 0$, $i=1,...,N$), where $L_{n_i}^{m_i} (x)$
is a Laguerre polynomial,
the ${\boldsymbol\eta}$-integration in Eq.~(\ref{wignerm})
can be carried out explicitly.
After that, using the orthogonality and completeness properties of
the Laguerre polynomials along the lines of Ref.~\cite{Casati08},
we can express the Wigner harmonics $W_{\bf m}({\bf I};t)$ in terms
of the matrix elements $\langle {\bf n}+{\bf m} |\hat{\rho}(t)|{\bf n}\rangle$
and finally obtain
\begin{equation}
\label{Wmrho}
\mathcal{W}_{\bf m}(t)=
\frac{\sum_{\bf n} |\langle {\bf n}+{\bf m}|
\hat{\rho}(t)|{\bf n}\rangle|^2}{{\sum_{m_1,...,m_N\ge 0}}
\sum_{\bf n} |\langle {\bf n}+{\bf m}|
\hat{\rho}(t)|{\bf n}\rangle|^2}.
\end{equation}

Finally, we point out that our approach remains valid also
for classical systems, provided the Wigner function is substituted by
the classical phase-space distribution function in the
$\alpha_k$-coordinates, with
$\alpha_k=\sqrt{I_k}e^{-i\theta_k}$, $\{I_k,\theta_k\}$ being
a set of action-angle variables for the unperturbed, integrable
Hamiltonian $H_0$.

\section{The Model}

\label{sec:model}

In order to investigate the working of our complexity measure,
we consider, as an illustrative example,
the Ising chain of $N$ spins in a tilted
magnetic field. The Hamiltonian reads
\begin{equation}
{\hat{H}}=J\sum_{i}\hat{\sigma}_{i}^{z}\hat{\sigma}_{i+1}^{z}
+\sum_{i}[h_{x}\hat{\sigma}^{x}_{i}+h_{z}\hat{\sigma}^{z}_{i}],\label{ising}
\end{equation}
where $J$ is the spin-spin coupling constant,
$\hat{\sigma}_i^\alpha$ are the Pauli operators for the $i$-th spin,
and $h_{x}$, $h_{z}$ are the field amplitudes along $x$ and $z$
directions, respectively. We set $\hbar=J=1$. This chain is in
general non-integrable, except for the two integrable limits
$h_{x}=0$ or $h_{z}=0$. The integrable model $h_z=0$ corresponds to
the Ising model in a transverse field and exhibits a quantum phase
transition at $J=h_{x}$~\cite{qpt}.

Using the Schwinger boson representation~\cite{Auerbach}, the above
spin Hamiltonian is mapped onto an interacting boson Hamiltonian.
Each spin operator $\hat{\sigma}_{i}$ at the site $i$  is replaced
by two Schwinger bosons, $\hat{a}_{i}$ and $\hat{b}_{i}$,
corresponding to spin up $ \{\uparrow\}$ and down $ \{\downarrow\}$.
The spin operators can be represented as follows:
\begin{eqnarray}\label{sch}
\hat{\sigma}_{i}^{z}=\hat{a}_{i}^{\dag}\hat{a}_{i}
    -\hat{b}_{i}^{\dag}\hat{b}_{i}, \nonumber \\
\hat{\sigma}_{i}^{+}=\hat{a}_{i}^{\dag}\hat{b}_{i},
\hspace{0.10in}\text{and}\hspace{0.10in}
\hat{\sigma}_{i}^{-}=\hat{b}_{i}^{\dag}\hat{a}_{i},
\end{eqnarray}
where $\hat{\sigma}^{\pm}=\frac{1}{2}(\hat{\sigma}^x\pm
i \hat{\sigma}^y)$. Since we have spin-1/2 particles, the
physical subspace is singled out by the constraints
$n_{ai}+n_{bi}=1$ ($i=1,...,N$),
where $n_{ai}$ and $n_{bi}$ denote the
number of up and down spins at site $i$
($\hat{n}_{ai}=\hat{a}_{i}^{\dag}\hat{a}_{i}$,
$\hat{n}_{bi}=\hat{b}_{i}^{\dag}\hat{b}_{i}$).
Now, Eq. (\ref{ising}) takes the form
\begin{eqnarray}
  {\hat{H}}=J\sum_{i=1}^{N-1}(\hat{a}_{i}^{\dag}\hat{a}_{i}
    -\hat{b}_{i}^{\dag}\hat{b}_{i})(\hat{a}_{i+1}^{\dag}\hat{a}_{i+1}
    -\hat{b}_{i+1}^{\dag}\hat{b}_{i+1}) \nonumber \\
    +\sum_{i=1}^{N}[h_{x}(\hat{a}_{i}^{\dag}\hat{b}_{i}+\hat{b}_{i}^{\dag}\hat{a}_{i})
    +h_{z}(\hat{a}_{i}^{\dag}\hat{a}_{i}
    -\hat{b}_{i}^{\dag}\hat{b}_{i})].\label{isingm}
\end{eqnarray}
As the Hamiltonian is now expressed in terms of a set of bosonic
creation-annihilation operators, it then follows that the above
explained phase-space approach can
be used to probe the dynamical complexity of the spin chain.

For a chain of $N$ spins, ${\bf n}$
and ${\bf m}$ in Eq.~(\ref{Wmrho}) are $2N$-dimensional vectors,
\begin{eqnarray}
{\bf n} &=& (n_{a1},n_{b1},n_{a2},n_{b2},...n_{aN},n_{bN}), \nonumber \\
 {\bf  m}&=& (m_{a1},m_{b1},m_{a2},m_{b2},...m_{aN},m_{bN}),
\end{eqnarray}
where the first subscript refers to the spin type and second refers
to the spin site.
The possible values of $n$'s are $0$ and $1$ with the constraint
$n_{ai}+n_{bi}=1$. Similarly, the possible values $m$'s can take are
$-1,0$ and $1$ with the constraint $m_{ai}+m_{bi}=0$ to remain on
the physical subspace.
Indeed, the following cases are possible:
(i) $m_{ai}=-1$, $m_{bi}=1$, corresponding to the transition of the
$i$-th spin from up to down,
(ii) $m_{ai}=1$, $m_{bi}=-1$ (transition of the
$i$-th spin from down to up), and
(iii) $m_{ai}=m_{bi}=0$ (no transition for the $i$-th spin).
Due to the trivial relation
$\mathcal{W}_{-{\bf m}}=\mathcal{W}_{\bf m}$, we
limit the summation (\ref{ent}) to $m_{ai}\ge 0$
(and, consequently, $m_{bi}\le 0$), for $i=1,...,N$,
in order to take into account independent terms only.
That is, for a chain of $N$ spins, only $2^{N}$
values of the $\mathcal{W}_{\bf m}$ are independent and are considered
for calculations. Note that the maximum possible value of the entropy
measure ${S}(t)$ is $N\ln2$. This value is reached
when maximum mixing occurs so that all the harmonics
are equally distributed.

\section{Phase-space Characterization of Complexity}

\label{sec:numerics}

\subsection{Initial Growth of $S(t)$}

In this section, we study in detail the time evolution of the
entropy ${S}(t)$ at small times. The initial state is
chosen to be a pure state with all spins pointing downward in the
$z$ direction, i.e.,
$|\Psi_{\text{in}}\rangle=|\downarrow\downarrow...\downarrow\rangle$.
From the definition of Wigner harmonics $\mathcal{W}_{\bf{m}}(t)$ in
Eq. (\ref{Wmrho}), it is clear that only the $\bf{m}$$=(0,0,...0)$
harmonics component is excited, the initial number of harmonics is equal to
unity and hence $S(t=0)=0$.

To relate the growth rate of the harmonics to the complexity of
the dynamics as in Ref.~\cite {Benenti09}, first we consider the short
time behavior of the entropy measure $S(t)$ when the system undergoes a transition
to quantum chaos as detected by a change of Poisson to Wigner
distribution in the
 level statistics. To that end, let us rewrite the Hamiltonian
${\hat{H}}$ as in Eq. (\ref{manyha}) with
\begin{equation}\label{xvary}
\hat{H}^{(0)}=J\sum_{i}\hat{\sigma}_{i}^{z}\hat{\sigma}_{i+1}^{z}
+\sum_{i}h_{z}\hat{\sigma}^{z}_{i};\hspace{0.20in}
\hat{H}^{(1)}=\sum_{i}h_{x}\hat{\sigma}^{x}_{i}.
\end{equation}
Here $\hat{H}^{(0)}$ is the integrable Hamiltonian and
 $\hat{H}^{(1)}$ represents the perturbation to the chain induced by an external transverse field.
 As the perturbation is
 increased, a transition to Wigner-type level statistics and hence quantum chaos occurs.
In particular, for
$h_{x}=h_{z}=J$
the system can be considered as fully chaotic~\cite{Prosen07}.

\begin{figure}
 \begin{center}
 \epsfig{file=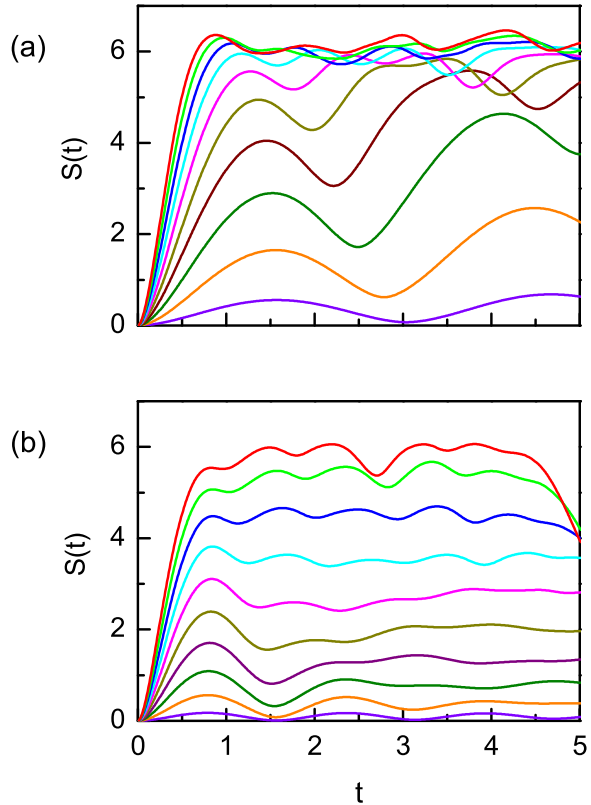,width=8.6cm}
 \end{center}
 \caption{(color online) Time dependence of the entropy measure ${S}(t)$ for a
 chain of $10$ spins with (a) non-integrable Hamiltonian as given in Eq.
 (\ref{xvary})
with a longitudinal field $h_{z}=1.0$ and (b) integrable Hamiltonian
as given in Eq. (\ref{xvaryin}). Curves from
 top to bottom correspond to transverse field $h_{x}=
 1.0$ to $0.1$ in decreasing steps of $0.1$. Note that during an initial time window $S(t)$ is clearly seen to grow linearly with time,
 implying the exponential growth of the number of harmonics in both (a) and (b), for a sufficiently large $h_{x}$. All the parameters
  mentioned here, and in the other figures, are dimensionless
(we set $\hbar=J=1$).}\label{xfieldt}
 \end{figure}

The initial state is an eigenstate of the unperturbed integrable
Hamiltonian $\hat{H}^{(0)}$. Hence with zero perturbation
$(h_{x}=0)$, there is no evolution and $S(t)$ remains zero at all
times. However, by adding a small perturbation to the system i.e., a
small transverse field $h_{x}$, the initial state is no longer an
eigenstate of the Hamiltonian $\hat{H}$ and transitions to many
other states occur. Then, besides the zeroth harmonics i.e.,
$\bf{m}$ $=(0,0,...0)$, higher harmonics are also excited and the
entropy $S(t)$ increases. Smaller the transverse field, less complex
is the dynamical evolution and therefore a lower value of the growth
rate for entropy ${S}(t)$ is obtained. This is clearly seen from our
results in Fig. \ref{xfieldt}(a). For instance, with $h_{x}=0.2$,
the chain is near the integrable regime and $S(t)$ is about $1.0$ at
time $t=1.0$. By contrast, in the chaotic regime with $h_{x}=0.9$,
${S}(t)$ increases to $6.3$. Interestingly, for cases with a
sufficiently strong perturbation, $S(t)$ is seen to increase
linearly with time within a time window ($t<0.6$). Such linear
increase of $S(t)$ implies an exponential growth of the number of
harmonics.

Consider then another situation for the
Ising chain, with the magnetic field applied in the transverse direction only:
\begin{equation}\label{xvaryin}
\hat{H}^{(0)}=J\sum_{i}\hat{\sigma}_{i}^{z}\hat{\sigma}_{i+1}^{z}
;\hspace{0.20in} \hat{H}^{(1)}=\sum_{i}h_{x}\hat{\sigma}^{x}_{i}.
\end{equation}
Note that this Hamiltonian is integrable for all values of the
perturbation $h_{x}$~\cite{lieb61}. On the basis of previous
findings in few-body problems~\cite{Gong03,Benenti09}, we might
expect a linear increase of the number of harmonics, corresponding
to a logarithmic growth of ${S}(t)$. On the contrary, as shown in
Fig. \ref{xfieldt}(b) the short time behavior of ${S}(t)$ is linear.
A comparison between Fig. \ref{xfieldt}(a) and Fig. \ref{xfieldt}(b)
clearly shows that one cannot distinguish between a chaotic and an
integrable many-body system by solely examining the initial growth
of the number of Wigner harmonics.

\begin{figure}
 \begin{center}
 \epsfig{file=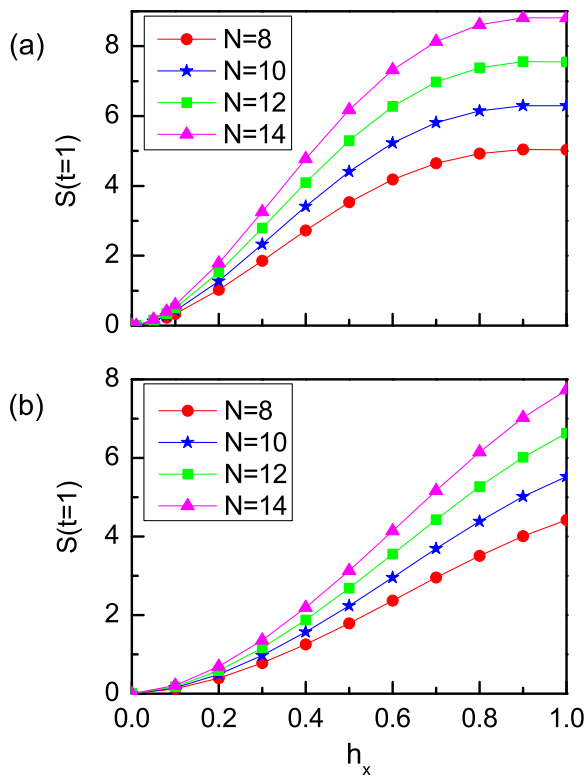,width=8.6cm}
 \end{center}
 \caption{(color online) Dependence of entropy $S(t)$ at time $t=1$ on the strength of the external perturbation $h_{x}$ for different chain length $N$ for (a)
 a non-integrable spin chain
 and (b) an integrable spin chain. For both cases and for sufficiently large $h_x$, it is seen that $S(t=1)$ scales linearly with $N$.}\label{xfieldN}
 \end{figure}

 To gain more insights, we next study how the initial linear growth rate of $S(t)$ depends on
 the number $N$ of spins in the chain, for both the non-integrable model (\ref{xvary}) and the integrable model
(\ref{xvaryin}).  In particular, we vary $N$ from $N=8$ to
 $N=14$.  Figure \ref{xfieldN}(a) shows the value of $S(t)$ at
 a fixed time $t=1$ as a function of $h_x$,  for the non-integrable model with four different values of $N$.
 Note that $S(t=1)$ can be understood as the average entropy production rate for $t\in[0,1]$. It can be observed that $S(t=1)$ scales with $N$ linearly. For example, for $h_x=1.0$, $S(t=1)$ increases by a constant value ($\approx 1$) as $N$ increases in steps of two. Interestingly, as shown in Fig. \ref{xfieldN}(b),
 exactly the same behavior is observed for the integrable model.  This further strengthens our early finding that the initial growth of the number of Wigner harmonics is qualitatively the same for non-integrable and integrable spin chains.  We must therefore seek
 an underlying mechanism to account for this
somewhat counter-intuitive behavior of many-body quantum systems.

\subsection{Wigner Harmonics and Entanglement}
A source of quantum complexity in many-body systems is the
entanglement due to the interaction between the different
constituent parts.  Recent studies
indicated that some measure of the entanglement entropy can also
grow linearly with time~\cite{Calabrese}.
We will therefore inquire whether or not the
lack of distinction between integrable and non-integrable models
shown above is related to the generation of multipartite
entanglement.

To quantify the extent of multipartite entanglement generated in a
spin chain, we adopt the multipartite entanglement measure used in
Ref.~\cite{Facchi06}.
 Specifically, the system under consideration is partitioned into two subsystems $A$
 and $B$, made up of $n_{A}$ and $n_{B}$ spins, respectively. The participation
 number $N_{AB}$, defined as the reciprocal of the purity of one of the two
 subsystems, i.e.,
 \begin{equation}\label{part}
    N_{AB}=\frac{1}{\text{Tr}[\hat{\rho}_{A}^{2}]},
\end{equation}
accounts for the bipartite entanglement between $A$ and $B$.
Here, $\hat{\rho}_{A}$ is the reduced density matrix of subsystem
$A$. The physical meaning of $N_{AB}$ is that it effectively counts the relevant terms in the Schmidt decomposition of the total wavefunction into the sum
of direct products of wavefunctions of the two subsystems.   The mean value of
  $N_{AB}$, averaged over all possible partitions, quantifies the degree of multipartite entanglement
  in the system, while its variance measures how well the entanglement
  is distributed. For systems of large size $N>>1$, the statistical
weight of unbalanced partitions becomes
  negligible~\cite{Facchi06}
and hence only balanced partitions are considered here.

\begin{figure}[h]
 \begin{center}
 \epsfig{file=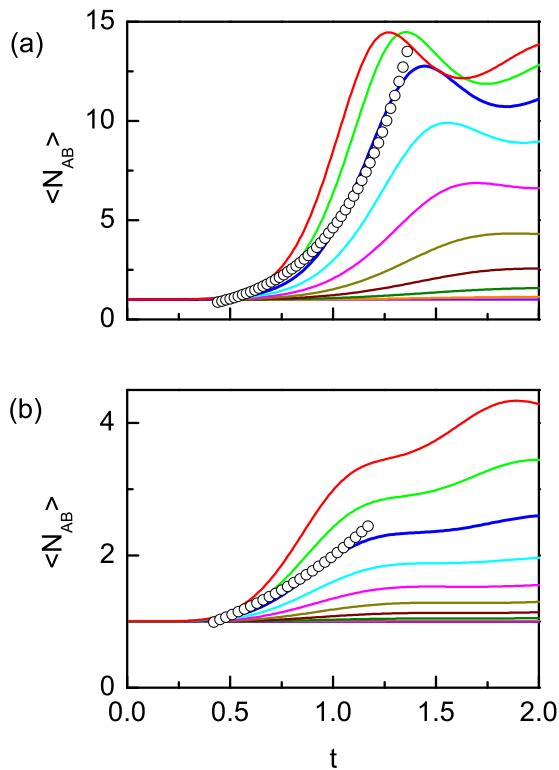,width=8.6cm}
 \end{center}
 \caption{(color online) Time dependence of average value of participation number $\langle N_{AB} \rangle$  calculated
 over all balanced bipartitions of the system for (a) non-integrable
 and (b) integrable model
 with parameters discussed in Fig. \ref{xfieldt}.
Curves from
 top to bottom correspond to transverse field $h_{x}=
 1.0$ to $0.1$ in decreasing steps of $0.1$.
Within a small time window,
 $\langle N_{AB} \rangle$ $\propto$ $e^{A t}$ for relatively large $h_x$,
 as shown by an exponential fit (circles) for $h_{x} = 0.8$ in both panels.
 }\label{gands}
 \end{figure}

In Fig.~\ref{gands}
we present the time dependence of the mean value of the participation number
 $\langle N_{AB}\rangle$, starting from the same initial state as before,
i.e.,
$|\Psi_{\text{in}}\rangle=|\downarrow\downarrow...\downarrow\rangle$,
for both integrable and non-integrable models.  The
initial state is not entangled and hence $\langle N_{AB}\rangle(t=0)$ is
given by its minimum value: unity. Entanglement is then generated
by the dynamical evolution of the spin chain and hence $\langle
N_{AB}\rangle$ increases. Remarkably, after a short time interval
$(t<0.4)$
and for a sufficiently large
value of $h_x$, $\langle N_{AB}\rangle$ reaches the saturation value
almost exponentially fast,
for both the integrable and non-integrable models.
Though the production of
entanglement is somewhat slower for the integrable chain as compared to the
non-integrable case, an exponential-like fast increase of $\langle N_{AB}\rangle$ is seen in both
situations. To visualize this more clearly, we plot an
exponential fit for $h_{x} = 0.8$ in both panels of Fig.
\ref{gands}.  These results indicate that we can ascribe the exponential growth of Wigner
harmonics to the fast entanglement generation in the chain.
\begin{figure}[h]
 \begin{center}
 \epsfig{file=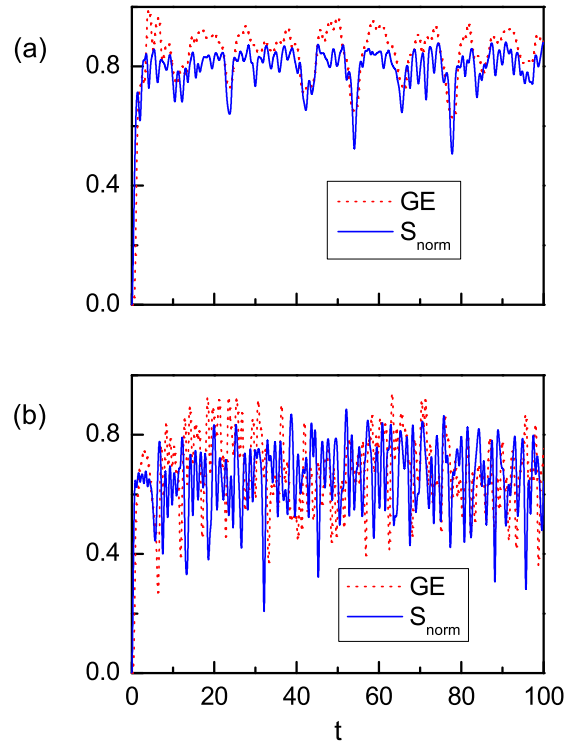,width=8.6cm}
 \end{center}
 \caption{(color online) Comparison of the dynamics of normalized entropy $S_{\text{norm}}$ and global entanglement
 $GE$ for a chain of $10$ spins with transverse field $h_{x}=0.8$. Panel (a) corresponds to the non-integrable model with longitudinal field $h_{z}=1.0$
 and panel (b) corresponds to the integrable model. Here $S_{\text{norm}}$
  is the entropy $S$ divided by its maximum value $N\ln 2$.  A close correspondence between the dynamics of two measures is evident.}
 \label{gent}
 \end{figure}

To better clarify this latter point,
we have compared  the time-dependence of
 $S(t)$ with the so-called
``global entanglement" (denoted $GE$)~\cite{Meyer02}.  In Ref. \cite{purity}, it was shown that $GE$ is
related to the averaged one-qubit purity, i.e.,
\begin{equation}\label{ge}
    GE=2\left(1-\frac{1}{N}\sum_{k=1}^{N}\text{Tr}[\hat{\rho}_{k}^{2}]\right),
\end{equation}
where $\hat{\rho}_{k}$ is the density matrix of the $k$-th spin after
tracing over all other spins in the system.
$GE$ is the average
bipartite entanglement over all possible bipartitions between a
single qubit and the rest of the system.
It is easy to see that $0\le GE \le 1$. Values of $GE$ close to one
indicate highly entangled many-body states. When a many-body  state
is not entangled, $GE$ equals to zero.

For the initial state
$|\Psi_{\text{in}}\rangle=|\downarrow\downarrow...\downarrow\rangle$,
we present in Fig. \ref{gent} a comparison between ${S}$ and $GE$.
To better visualize their similarities, we plot a normalized
(to unity) entropy ${S}_{\text{norm}}={S}/N\ln 2$.
It is clearly observed that these two quantities show a high degree of resemblance in their time dependence.
Their oscillating patterns are quite close and in some regimes they are almost
on top of each other.
 The similarity between ${S}_{\text{norm}}$ and $GE$ constitutes strong evidence that our entropy measure ${S}(t)$, though originated from considerations of phase-space complexity, also reflects the degree of multipartite entanglement
in many-body systems.

\begin{figure}[h]
 \begin{center}
 \epsfig{file=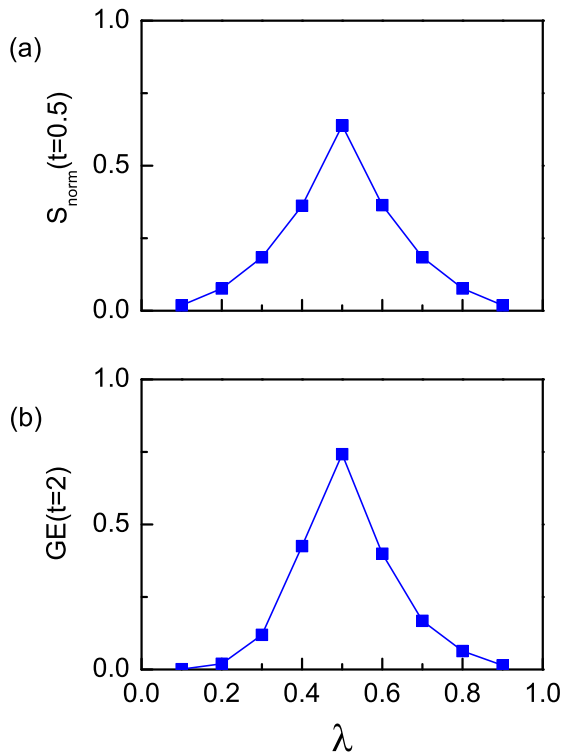,width=8.6cm}
 \end{center}
 \caption{(color online) Panel (a) shows $S(t=0.5)$ normalized by a factor of $N\ln 2$ as a function of the transverse field $h_{x}$
 for a transverse Ising chain of $10$
 spins. Panel (b) shows the parallel results for $GE$ at $t=2.0$.  Here $\lambda\equiv
 h_{x}/(J+h_{x})$. Similar to $GE$, the complexity measure $S$ is seen to peak clearly
at the critical point $\lambda=1/2$.
 }\label{qpt}
 \end{figure}

Though, in general, $GE$ may not distinguish between different classes
of multipartite entangled states, it is an indicator of the critical
point of, for instance, the quantum phase transition for the Ising
chain in a transverse magnetic field~\cite{Oliveira}. Therefore, it
is also interesting to investigate the behavior of $S(t)$ in the
neighborhood of a quantum critical point which, for the transverse Ising
chain in Eq. (\ref{xvaryin}) with coupling strength $J=1$, is
at $h_{x}=1$.  In Fig. \ref{qpt}, we show
the behavior of ${S}(t=0.5)$ as well as $GE(t=2)$, as a function
of $\lambda\equiv h_{x}/(J+h_{x})$. Note that ${S}$ and $GE$ are
plotted at different times because of their different saturation
times. In addition, in our calculations of $S$ on the large-field
side ($\lambda>1/2$) the $x$ axis is used as the quantization axis
of the basis states: The magnetic field term is dominant and
correspondingly we define
$\hat{H}^{(0)}=\sum_{i}h_{x}\hat{\sigma}^{x}_{i}$,
$\hat{H}^{(1)}=J\sum_{i}\hat{\sigma}_{i}^{z}\hat{\sigma}_{i+1}^{z}$,
and Schwinger bosons such that
$\hat{\sigma}_{i}^{x}=\hat{a}_{i}^{\dag}\hat{a}_{i}-\hat{b}_{i}^{\dag}\hat{b}_{i}$.
It is quite natural to consider as preferential basis the one
associated with the dominant term in the Hamitonian: the $z$-basis
when $\lambda\to 0$ and the $x$-basis when $\lambda\to 1$, and quantum
phase transition corresponds to the switching from one preferential
basis to the other.
Consistent with the expectation that the quantum phase transition
occurs at $\lambda=1/2$,  Fig.~\ref{qpt}(a) shows that
${S}_{\text{norm}}(t=0.5)$, a measure of the growth rate of the
number of Wigner harmonics, exhibits a sharp peak at $\lambda=1/2$.
The $\lambda$-dependence of $GE$ shown in Fig.~\ref{qpt}(b) is
analogous to what we observe in Fig.~\ref{qpt}(a). This further
demonstrates the close connection between our complexity measure
$S(t)$ and the global entanglement $GE$ and, in particular, the role
of many-body entanglement in the initial growth of $S(t)$.

 Two additional aspects of $S(t)$ are in order.  First, if we stick to the $z$ axis as the quantization axis of the basis states,
 then it is found that right after the critical point, $S(t)$ (if averaged over a time window to remove fluctuations) will show clear saturation behavior, which is in contrast to the monotonous increase of $S(t)$ before the critical point. Second, if we switch the quantization axis from $x$ to $z$ at other values of $\lambda$, then the value of $S(t=0.5)$ jumps discontinuously due to the change of the basis states.  These additional results further suggest that the critical point for quantum phase transitions can be detected by $S(t)$.

Note that
a different phase-space measure~\cite{sugita} has been used in the
literature to detect quantum phase transitions~\cite{ingold}.
However, such measure accounts for the extent at which the phase
space is covered by the Husimi distribution and therefore it does
not appear clear how to extend it to a suitable complexity measure
for mixed states. In contrast, our measure which is based on the
richness of the phase space structure rather than on phase-space
coverage can be used for both pure and mixed quantum states. For
instance, it could be used also to investigate thermal phase
transitions.

\subsection{Wigner Harmonics, Chaos, and Thermalization}

Our results so far indicate that due to the dynamically generated
many-body entanglement, the initial time-dependence of $S(t)$ does
not reflect the peculiarity of quantum chaos in many-body quantum
systems: it behaves similarly in integrable and non-integrable
models.  Note that this does not contradict with previous findings
regarding rapid bipartite entanglement generation in classically chaotic systems with two degrees of freedom \cite{bipartite}.
Indeed, in systems with two degrees of freedom, the Hilbert space is only a product of two subspaces and the quantum dynamics can only generate bi-partite entanglement. The rate of entanglement growth within such a fixed product of two subspaces is connected with the underlying classical dynamics.
By contrast, in a many-body system such as our model used here the dynamics emanating from a local initial state is seen to
explore more and more the tensor-product structure
of the total Hilbert space and hence entangle more and more degrees of freedom during the time evolution.  Our entropy measure $S(t)$ then is indicative of the growth rate of the number of degrees of freedom that have been entangled by the dynamics (a property absent in few-body systems).

Since the short-time behavior of $S(t)$ is seen to be unrelated to quantum chaos, we now examine the manifestation of
quantum chaos in the long-time behavior of
 $S(t)$. This can be justified because after all, the peculiar spectral statistics of a quantum chaotic many-body system
reflects the long-time properties of the system.

\begin{figure}[h]
 \begin{center}
 \epsfig{file=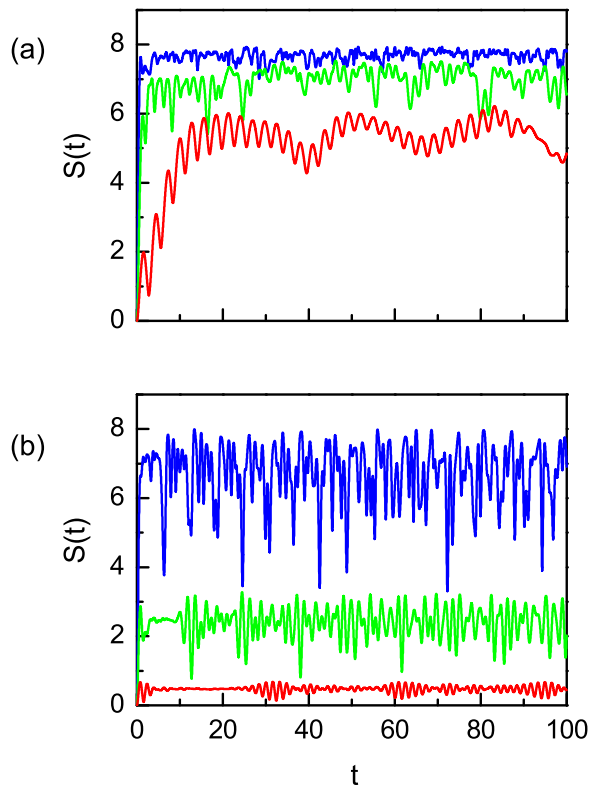,width=8.6cm}
 \end{center}
 \caption{(color online) Time evolution of entropy ${S}(t)$ for a chain of $12$
 spins with (a) non-integrable
 Hamiltonian in Eq. (\ref{xvary}) for $h_{z}=1.0$ and (b) integrable Hamiltonian in Eq.~(\ref{xvaryin}). Curves from top to bottom
 corresponds to transverse field $h_{x}$ = 1.0, 0.5 and 0.2. Note that the long term dynamics of the entropy in the two panels is
 qualitatively different.}
 \label{long}
 \end{figure}

\begin{figure}[h]
 \begin{center}
 \epsfig{file=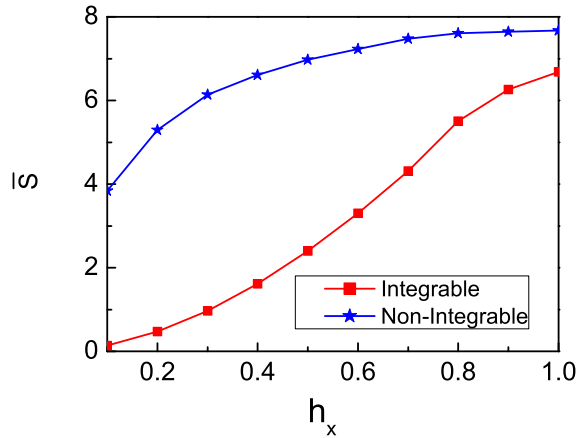,width=8.6cm}
 \end{center}
 \caption{(color online) Time averaged entropy, denoted $\bar{S}$, vs the strength of the transverse field $h_x$, for a chain of $12$ spins.
The non-integrable model corresponds to the Hamiltonian in Eq.
(\ref{xvary}) with $h_{z}=1.0$; the integrable model corresponds to
the Hamiltonian in Eq. (\ref{xvaryin}).  The difference in $\bar{S}$
between integrable and non-integrable dynamics decreases with
increasing $h_x$.} \label{avg}
 \end{figure}

The time-dependence of $S(t)$ for a non-integrable model with the
Hamiltonian given in Eq. (\ref{xvary}) is shown in
Fig.~\ref{long}(a). The parallel result for an integrable model
defined by Eq. (\ref{xvaryin}) is shown in Fig.~\ref{long}(b). In
both cases, the time scale under study is now 20 times longer than
that used in Fig.~1.  It is seen that, in both panels (a) and (b),
$S(t)$ initially quickly increases and then displays saturation with
rich oscillating behavior. The saturation plateau of $S(t)$
increases as
 the value of the transverse field increases.  Qualitatively, the saturation plateau
can be attributed to an effective dimension of the Hilbert space
that can be explored for a particular strength of the transverse
field. To quantitatively describe this observation,  we calculate
the average value of the entropy from time $t_1=5$ to $t_2=100$
as $\bar{{S}}=\frac{1}{\tau}\int_{t_1}^{t_2} dt {S}(t)$, where
$\tau=t_2-t_1$.
The values of $t_1$ and $t_2$ are chosen so that the saturation
plateau is reached before $t_1$ and $t_2$ is large enough
to allow averaging over many oscillations of $S(t)$ in the time
interval between $t_1$ and $t_2$. We have checked that
other such choices of $t_1$ and $t_2$ do not affect any of
our observations reported below.
In Fig. \ref{avg}, we plot $\bar{{S}}$ as a
function of the transverse field, for the integrable and
non-integrable models considered in Fig.~\ref{long}. It is seen
that, for small values of $h_x$, there is a difference between
integrable and non-integrable dynamics.  However, as the strength of
the transverse field increases, this difference reduces. This is
somewhat expected due to the above-discussed many-body entanglement generation.

\begin{figure}[h]
 \begin{center}
 \epsfig{file=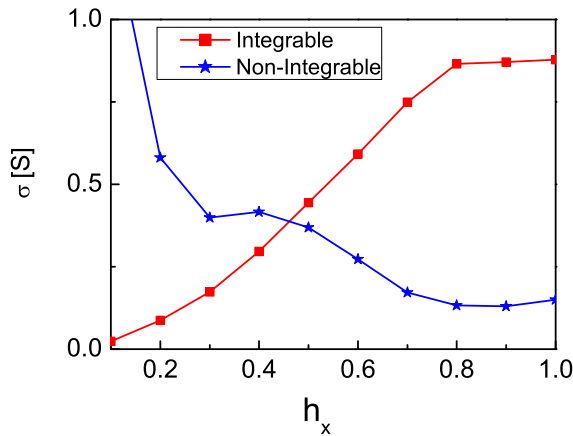,width=8.6cm}
 \end{center}
 \caption{(color online) Standard deviation $\sigma[S]$
in the entropy ${S}(t)$ vs the strength of the transverse field $h_{x}$ for a chain of $12$ spins. Here the integrable and
 non-integrable models are the same as in Fig. \ref{avg}.  It is observed that as $h_x$ increases,
 the standard deviation
generally decreases in the non-integrable model, but increases
 in the integrable model. For large values of $h_x$,
$\sigma[S]$ for a non-integrable chain is much smaller than that for an integrable chain.}
 \label{varis}
 \end{figure}

In order to distinguish between integrable and non-integrable cases,  we are thus forced to look into the oscillating behavior (rather
than the average behavior) of $S(t)$.  Indeed, from Fig.~\ref{long}
one observes that in the non-integrable case, the oscillation
amplitude of $S(t)$
clearly decreases with the value of $h_x$. The oscillation pattern
also becomes erratic as the system gets closer to the chaotic regime. By
contrast, in the integrable case the opposite trend is observed.
Regular and strong quantum revivals in $S(t)$ become more apparent
as $h_x$ increases. To quantitatively describe this clear
difference,  we calculate the standard deviation
of $S(t)$ around the mean
value $\bar{S}$,
\begin{equation}\label{var}
    \sigma[S]=\sqrt{\frac{1}{\tau}\int_{t_1}^{t_2}  d t [{S}(t)-{\bar{S}}]^{2}}.
\end{equation}
The results are shown in Fig.~\ref{varis}. With increasing
perturbation, the standard deviation
$\sigma[S]$ increases and then saturates in the
integrable model, so that the relative size
$\sigma[S]/\bar{S}$ of fluctuations remains nearly constant.
However, in the non-integrable model, the
standard deviation $\sigma[S]$ and, more markedly, $\sigma[S]/\bar{S}$
decrease during the regular-to-chaotic crossover
(The same qualitative behavior is obtained when the number of the
spins $N$ in the chain is varied).  For $N=12$, $\sigma[S]$ in the
integrable model with $h_{x}=1$ is around $0.878$. By sharp
contrast, in the non-integrable case with $h_{x}=1$,
$\sigma[S]=0.1495$, which is smaller than the first case by more than $5$
times. This is a dramatic difference considering that the total
number of spins in the chain is only $12$.

\begin{figure}[h]
 \begin{center}
 \epsfig{file=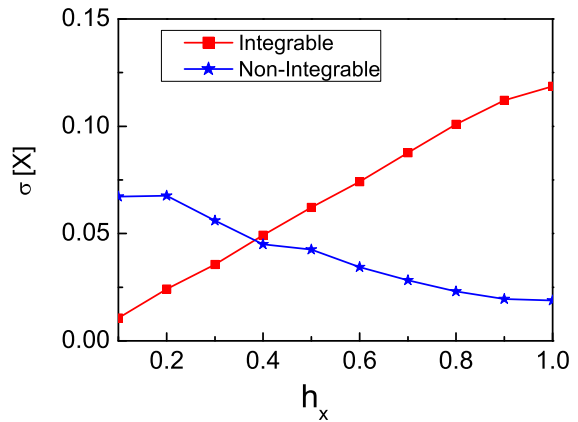,width=8.6cm}
 \end{center}
 \caption{(color online) Standard deviation of the $x$-polarization expectation value as a function of transverse field $h_{x}$ for a chain of $12$ spins. Here the integrable and
 non-integrable models are the same as in Fig. \ref{avg}. Similar to the case of entropy $S(t)$, with increasing perturbation, the standard deviation
$\sigma[X]$ decreases in the non-integrable model and increases
 in the integrable model. }
 \label{varsi}
 \end{figure}

The large value of the standard deviation for the integrable model can be
accounted for by the lack of thermalization.
Indeed, our expectation is that
the onset of chaos leads to internal dynamical
thermalization~\cite{thermalization}, so that a statistical description is
possible even though we have a closed, finite Hamiltonian system.
Since the density of many-body energy levels grows exponentially
with the number of particles, even a weak interaction between particles
typically leads to a strong mixing on noninteracting many-body states,
thus resulting in chaotic eigenstates. That is to say, the components
of such eigenstates can be treated as random variables and therefore
statistical methods can be applied to the description of local
observables, in spite of the fact that close systems are under consideration.
In such situation, fluctuations of
the expectation values of local observables are small.
On the contrary, in the integrable regime the lack of thermalization
allows large fluctuations.
To verify the above expectations, we have considered
the Pauli operator $\hat{\sigma}^x$
(note that, due to translational invariance of model
and initial condition,
$\langle \hat{\sigma}_i^x \rangle(t)$ is independent of $i$
at any time $t$).
Observing the long term dynamics of the chain, we compute
in the same way as in Eq.~(\ref{var}) for $S(t)$
the standard deviation $\sigma[X]$ of
the $x$-polarization expectation value
$X(t)\equiv \langle \hat{\sigma}^x \rangle(t)$.
Our results show that in the integrable model, the
standard deviation increases with the transverse field whereas
for the non-integrable case, the standard deviation decreases.
This behavior, illustrated in Fig. \ref{varsi}, is qualitatively
similar to the behavior of $S(t)$ shown in Fig. \ref{varis}.
This shows that our complexity measure is related to the
thermalization properties of the system.

\section{Conclusions}

\label{sec:conclusions}

In this paper,
we propose an entropy measure $S(t)$ for many-body quantum-dynamical
complexity, by extending the Wigner harmonics measure
introduced in~\cite{Gong03,Casati08,Benenti09} for single-particle quantum
dynamics.
The effectiveness of this measure is illustrated
in the example of
the Ising chain in a homogeneous tilted magnetic field.
The Wigner harmonics entropy $S(t)$ exhibits an initial
linear growth in both integrable and chaotic regimes, until
saturation occurs due to the finite size of the Hilbert space.
Therefore, in both integrable and chaotic regimes the number
of harmonics of the Wigner function grows exponentially with time.
In classical dynamics, an exponential growth of the number of
harmonics of the classical phase-space distribution function
implies chaotic dynamics. Therefore, the observed exponential
growth of Wigner harmonics in the many-body quantum integrable regime
must be attributed to a source of complexity absent in classical
dynamics, that is, entanglement. We have numerically demonstrated
the close connection between our complexity measure $S(t)$ and
multipartite entanglement, thus providing evidence that the
initial linear growth of $S(t)$ has to be ascribed to
multipartite entanglement generation.
The Wigner harmonics measure $S(t)$ can also distinguish
between integrable and chaotic many-body systems,
by means of the size of long term fluctuations, which
are smaller in the chaotic regime where a statistical
description of the system is legitimate and the relative
size of fluctuations drops when the system size increases.

The main advantage of the phase-space approach to complexity
resides in its generality. At the classical limit, the harmonics
of the phase-space distribution function reproduce the well-known notion of
complexity based on local exponential instability~\cite{Gong03}:
the number of harmonics grows linearly for integrable systems
and exponentially for chaotic systems.
In single-particle quantum mechanics, an exponential growth of
the number of harmonics is possible only up to the Ehrenfest
time scale, after which the growth is at most linear~\cite{Casati08}.
Furthermore, the number of harmonics of the Wigner function can
be used to detect, in the time domain, the crossover from
integrability to chaos~\cite{Benenti09}. For quantum many-body systems,
the Wigner harmonics entropy measure $S(t)$ proposed in this
paper signals the generation of multipartite entanglement
and can be used to detect quantum phase transitions.
In relation to other measures of complexity
based on the efficiency of the best classical simulations of
quantum systems~\cite{Prosen07}, our approach has the advantage
that it does not rely on a specific computational method
like the time-dependent density-matrix renormalization group.
Finally, we point out that, in contrast to other quantum phase-space
approaches based on the moments of the Husimi function~\cite{sugita,ingold},
our complexity measure works equally well for either pure or
mixed quantum states. Therefore, our measure could be studied
in relation to mixed-state entanglement. This would be particularly
interesting as mixed-state entanglement is at present not well
understood and is the focus of ongoing research.

        \end{document}